\documentstyle[osa,manuscript,psfig]{revtex} 

\begin{document}
\titlepage

\title{Inelastic diffraction and meson radii}

\author{Meng Ta-chung, R. Rittel, K. Tabelow, 
and Zhang Yang\footnote{{\em Present address:} Chinese Academy of
Sciences, Institute of Theoretical Physics, POB 2735, Beijing 100080,
China} \protect\\
{\em Institut f\"ur Theoretische Physik, FU-Berlin,
14195 Berlin, Germany}\protect\\
{\small e-mail: meng@physik.fu-berlin.de;
                rittel@physik.fu-berlin.de;
                tabelow@physik.fu-berlin.de;
		zhangy@itp.ac.cn} }

\maketitle

\begin{abstract}
It is shown that meson radii
 can be extracted
from diffractive 
scattering experiments of the type: 
$m+p\rightarrow m + X$, where
$p$ stands for the target proton, and $m$ for the incident charged meson
$\pi^+$, $\pi^-$, $K^+$, or $K^-$. The 
basis of the method used to perform this extraction is
the QCD-based
SOC-picture for inelastic diffraction in high-energy hadron-hadron and 
lepton-hadron collisions proposed in a recent Letter. The obtained
results for pion and kaon charge radii are 
compared with those determined in meson form factor 
measurements.  
\end{abstract}


Ever since it is known that not only nucleons, but also pions and kaons, 
are spatially extended objects, determination of meson radii has been 
of considerable interest --- both experimentally and theoretically
(see e.g. Refs.[\ref{electro1}-\ref{direct7}]
and papers cited therein).
Independent of the fact whether the collected
information on meson radii is obtained from
electroproduction \cite{electro1,electro2,electro3}, 
from $e^+e^-$-annihilation \cite{collider1,collider2},
or from direct meson-electron scattering 
\cite{direct1,direct2,direct3,direct4,direct5,direct6,direct7} experiments, 
such information have always been extracted from meson form
factors --- based on Feynman graphs (or subgraphs) in which the meson
is scattered elastically in an electromagnetic field.
Can we determine the spatial extension of a meson without any
knowledge of its electromagnetic form factor?

In this Brief Report, we show
that meson radii can be extracted from inclusive
differential cross-section data for diffractive scattering processes
of the following type: $m + p \rightarrow m +X$ where $p$ stands for
the proton target and $m$ is the incident meson:
$\pi^+, \pi^-, K^+,$ or $K^-$. The method used for this extraction is
a QCD-based SOC-picture for
inelastic diffraction in high-energy hadron-hadron and lepton-hadron
collisions proposed in a recent Letter \cite{letter}, in which the 
following are pointed out: 

First, systems of interacting soft gluons which play the dominating
role in diffractive scattering processes are extremely complex --- so
complex that concepts and methods developed for the description of 
Complex Systems, in particular those in connection with the existence
of Self-Organized Criticality
\cite{BTWoriginal,BTWcontinue1,BTWcontinue2}, 
are needed in understanding
the observed phenomena. The applicability of such concepts and methods 
are justified, and the existence of self-organized criticality
(SOC) can be expected, because of the following: (A) The characteristic 
properties of the gluons (especially the direct gluon-gluon coupling
prescribed by the QCD-Lagrangian, the confinement, and the
non-conservation of gluon-numbers) strongly suggest that systems of
interacting soft gluons are open, dynamical, non-equilibrium, complex
systems with many degrees of freedom. (B) It is this kind of systems
which has been intensively studied in Sciences of Complex Systems,
and it is in this kind of systems where the 
fingerprints of SOC have been found. (C) The existence of
SOC-fingerprints can be, and have been, checked \cite{letter,longSOC}
in systems of interacting soft gluons by examing high-energy
scattering processes in which such gluons dominate. It is observed in
particular that the  ``colorless objects'' which play the
dominating role in inelastic diffractive scattering processes can be 
considered as color-singlet gluon-clusters in form of BTW-avalanches 
\cite{BTWoriginal,BTWcontinue1,BTWcontinue2} due to SOC in systems of
interacting soft gluons. To be more precise:
Such colorless gluon-clusters 
are in general partly inside and partly outside
the proton. That is, they form a kind of
``cluster cloud''\cite{letter,longSOC}. 
Since the average binding energy between color-singlet aggregates is
of Van der Waals type, which implies that this kind of binding energy
is negligible small  compared with
the corresponding one between colored objects, it is
expected that, even at  relatively small values of momentum-transfer
($|t|$$<$$1\,\mbox{GeV}^2$, say), the struck colorless clusters can
be absorbed by the beam-particle, and thus ``be carried away''
by the latter.

Second, 
optical-geometrical concepts and methods can be used to examine the
space-time properties of the above-mentioned colorless objects. 
In fact,
having the well-known phenomena associated with Frauenhofer's
diffraction and the properties of de Broglie's
matter waves in mind, the projectiles $P$
($\gamma^\star$, $\gamma$, $\overline{p}$, or $p$) 
in a scattering process 
\begin{eqnarray}\label{eq3}
P+T&\rightarrow & X+T \, ,
\end{eqnarray}
where $T$ stands for the target proton ($p$),
$P$ stands for the projectile which can be an antiproton ($\bar{p}$),
a proton ($p$), a virtual photon ($\gamma^\star$), or a real photon
($\gamma$),
that is
\begin{eqnarray}\label{eq4}
& &\left\{
\begin{array}{l}
P=\gamma^\star, \gamma, p, \mbox{\ or \ } \bar{p},\\
T=p.
\end{array}\right.
\end{eqnarray}
can be viewed as high-frequency waves passing through a medium, 
where the medium is the above-mentioned  
``cloud'' of color-singlet gluon-clusters.

In fact, as we have already pointed out in Ref.[\ref{letter}],
the proposed picture for
{\em inelastic} diffraction for high-energy collisions has two
ingredients which are fundamentally different from those in the
conventional picture for {\em elastic} diffraction: 
(i) ``The scatterer'' in
the proposed picture for {\em inelastic} diffraction is {\em not} a
static object. It is a Complex System which depends on space and time!
(ii) In the proposed picture for {\em inelastic} diffraction, the
wave-vector of the outgoing de Broglie wave ($X$) differs from the
incoming one ($P$), not only in direction, but also in magnitude. That 
is, the frequency and the longitudinal component of the wave-vector of 
the outgoing wave ($X$) can differ very much from their counterparts
in the incoming wave ($P$). This is why, and how, relatively large
energy-momentum transfer in passing the medium can result in large
invariant mass $M_x$ for $X$ which is described as outgoing waves in
such {\em inelastic} diffraction processes. 

Based on the proposed picture\cite{letter}
simple analytical formulae for the
differential cross-sections $d\sigma/dt$ and $d^2\sigma/dt d(M_x^2/s)$ can
be, and have been,
derived \cite{letter,longSOC} for inelastic diffractive scattering 
processes.They are
\begin{eqnarray}
\label{eq5}
\frac{d\sigma}{dt} &=& C\exp (-2a\sqrt{|t|}) \\
\label{eq6}
\frac{1}{\pi}\frac{d^2\sigma}{dx_P \, dt} &=& N x_P^{-2/3}\exp (-2a\sqrt{|t|}),
\end{eqnarray} 
where $C$ and $N$ are unknown (normalization) constants, which are
obviously related to each other, 
$x_P=M_x^2/s$ ($\sqrt{s}$ is the total cms energy, and $M_x$ is the
invariant mass of the unidentified hadronic system $X$), $t$ is the
invariant momentum transfer (which is, for $x_P \ll 1$, approximately
$-p^2_\bot$, where $p_\bot$ is the transverse momentum of the
scattered particle). The constant $a$ is, because of SOC, QCD, and
confinement, directly related (see Ref.[\ref{letter}] and/or
Ref.[\ref{longSOC}] for details) to the proton's charge 
radius $r_p \equiv <r_p^2>^{1/2}$:
\begin{equation}
\label{eq7}
a=\sqrt{\frac{3}{5}} r_p\mbox{.}
\end{equation} 

While the derivation of Eqs.(\ref{eq5}), (\ref{eq6}), and (\ref{eq7})
presented in [\ref{letter}] and [\ref{longSOC}] will not be
repeated here, we do wish to emphasize the following:
The reason why the charge radius is used in these equations as a
measure of proton's confinement region is also closely related with
the fact that gluons carry color. 
To be more precise, the fact that gluons carry color,
not only implies that the spatial location where the
(colored and colorless) gluon-clusters in form of BTW-avalanches are
initiated  
has to be inside the confinement region of the proton,
but also implies the following: 
In accordance with QCD, the chance for colored objects to
create seaquark-antiseaquark ($q_s \bar{q_s}$) pairs is larger than that 
for a corresponding colorless objects to do so. This means, the created
$q_s \bar{q_s}$-pairs are mainly due to the colored objects
--- all of which (including the valence quarks and the gluons)
have to remain in the confinement region of the target proton.
Having in mind, that the electric charge of $q_s$ and that of
$\bar{q_s}$ can be detected by an incident (real or virtual) 
photon, we see that 
the region in which electric charge can be detected,
and the region in which BTW-avalanches in gluon systems can be
initiated, are (at least approximately) {\em the same}.
This is the reason why charge radius of the 
proton plays the important role in Refs.[\ref{letter}] and [\ref{longSOC}],
and this is also why useful information on meson radii can
be expected by looking at the corresponding single diffractive
scattering processes. 

In order to extract meson radii from data for inelastic diffraction in 
high-energy collision processes, we now examined single diffractive 
scattering processes of the type 
\begin{eqnarray}\label{eq8}
P+T&\rightarrow & P+X \, ,
\end{eqnarray}
where $P$ is a charged meson and $T$ a proton, that is
\begin{eqnarray}\label{eq9}
& &\left\{
\begin{array}{l}
P=m=\pi^-,\pi^+,K^-, \mbox{\ or \ } K^+,\\
T=p \, .
\end{array}\right.
\end{eqnarray}
Having the proposed \cite{letter}
QCD-based SOC-picture for inelastic diffraction in high-energy
scattering processes in mind, the apparent symmetry between 
the reactions mentioned in Eqs.(\ref{eq3}) and
(\ref{eq4}) and those mentioned in Eqs.(\ref{eq8}) and (\ref{eq9})
show that they are closely related to one other, as we can explicitly
see in Fig.1.

In fact, it follows immediately from Fig.1, that
the differential cross-sections $d\sigma/dt$ and $d^2\sigma/dt dx_P$ 
in the corresponding
kinematical regions should have the same kind of behaviors
as those shown in Eqs.(\ref{eq5}), (\ref{eq6}) and (\ref{eq7}) 
with the following modification: 
The proton radius in Eq.(\ref{eq7}) should be replaced by
the corresponding meson radius namely by $r_\pi$ for pions, and by 
$r_K$ for kaons. It is because, in this case, the struck color-singlet
gluon-cluster in form of a BTW-avalanche is initiated {\em in the meson},
while the proton in the reactions mentioned in (\ref{eq8}) and
(\ref{eq9}) appears
in form of a de Broglie wave which undergoes Frauenhofer diffraction.
This means, based on the picture proposed in Ref.[\ref{letter}],
it should be possible to determine the radius of the charged meson
from the inclusive cross section data \cite{diff-data} obtained in 
the corresponding {\em inelastic diffractive scattering experiments}.

In order to put this idea into practice, we need to compare these
formula with the data \cite{diff-data} 
and thus extract the ``slope parameter'' if the
agreement is reasonable. Here we have to take the following 
facts into account: First, according to Eqs.(\ref{eq5}) and
(\ref{eq6}), it is only the dependence of the differential cross
section on $p_\bot$ (recall that $p_\bot$ is approximately
$\sqrt{|t|}$ for $x_P\ll 1$) 
which is directly associated with the meson radius.
Second, there seems to be no
integrated data for $d\sigma /dt$ (or $d\sigma /dp_\bot$) 
for the considered reactions.
Based on these facts we performed a fit to the data \cite{diff-data} 
for the double differential cross setion for 
$m+p\rightarrow m+\mbox{anything}$ with $m=\pi^+, \pi^-, K^+, K^-$,
where the $p_\bot$-dependence of the differential cross sections is
given by (\ref{eq6}) for every given value of $x_P=M_x^2/s$. In these 
fits, the charges of the pions and those for the kaons are ignored
in order to have better statistics. 
The results of these fits together with the corresponding values for
the meson radius $r_m$ are shown in the Figs. 2 and 3, respectivly.
The value for the radius is obtained by averaging over the values for $r_m$ 
at different $x_P$ weighted by
the number of points (minus one) in each box. 
The results are
\begin{eqnarray}\label{eq10}
r_{\pi^\pm}&=& 0.62 \pm 0.04 \,\, \mbox{fm} ,\\\label{eq11}
r_{K^\pm}&=& 0.53 \pm 0.03 \,\, \mbox{fm}.
\end{eqnarray}
A comparison with the results obtained by using
other methods is shown in Figs. 4(a) and 4(b).

In addition, we recall that, in order to minimize systematic errors, 
the difference of the squared
radii of pions and kaons $<r^2_\pi>-<r^2_K>$ has been measured and 
given in Refs.[\ref{direct3}] and [\ref{direct7}]. 
The experimental values are
$0.16 \pm 0.06 \mbox{ fm}^2$ in \cite{direct3}, and
$0.10 \pm 0.045 \mbox{ fm}^2$  in \cite{direct7}, respectively.
Our result for this difference extracted from the data given in
Ref.[\ref{diff-data}] is: 
\begin{equation}
\label{eq12}
<r^2_\pi>-<r^2_K>=0.10 \pm 0.07 \mbox{ fm}^2 
\end{equation}
A graphical comparison is shown in Fig. 4(c).

In conclusion, we see that, by applying the physical picture
proposed in Ref.[\ref{letter}] to single diffractive
scattering processes, in which charged mesons are used as projectile,
and produced hadrons are found in the fragmentation region of the
proton target, meson radii can be extracted from
{\em inelastic diffractive scattering data} \cite{diff-data}. The
obtained results for pion and kaon charge radii are in good agreement 
with those obtained from meson form factor measurements (see
Refs.[\ref{electro1}-\ref{direct7}] and the papers cited therein.). 
Since this application can be considered as a
crucial test of the usefulness of concepts and methods developed for
Complex Systems [in particular those in connection with the existence
of Self-Organized 
Criticality], in understanding the mechanisms of reactions where
interacting gluons play a dominating role, the results presented in
this Brief Report can be considered as further confirmation of
the usefulness of this approach.

The authors thank Cai Xu for helpful discussions, and FNK der FU for
financial support. Y. Zhang also thanks Alexander von Humboldt
Stiftung for the fellowship granted to him.

\begin{figure*}
\psfig{figure=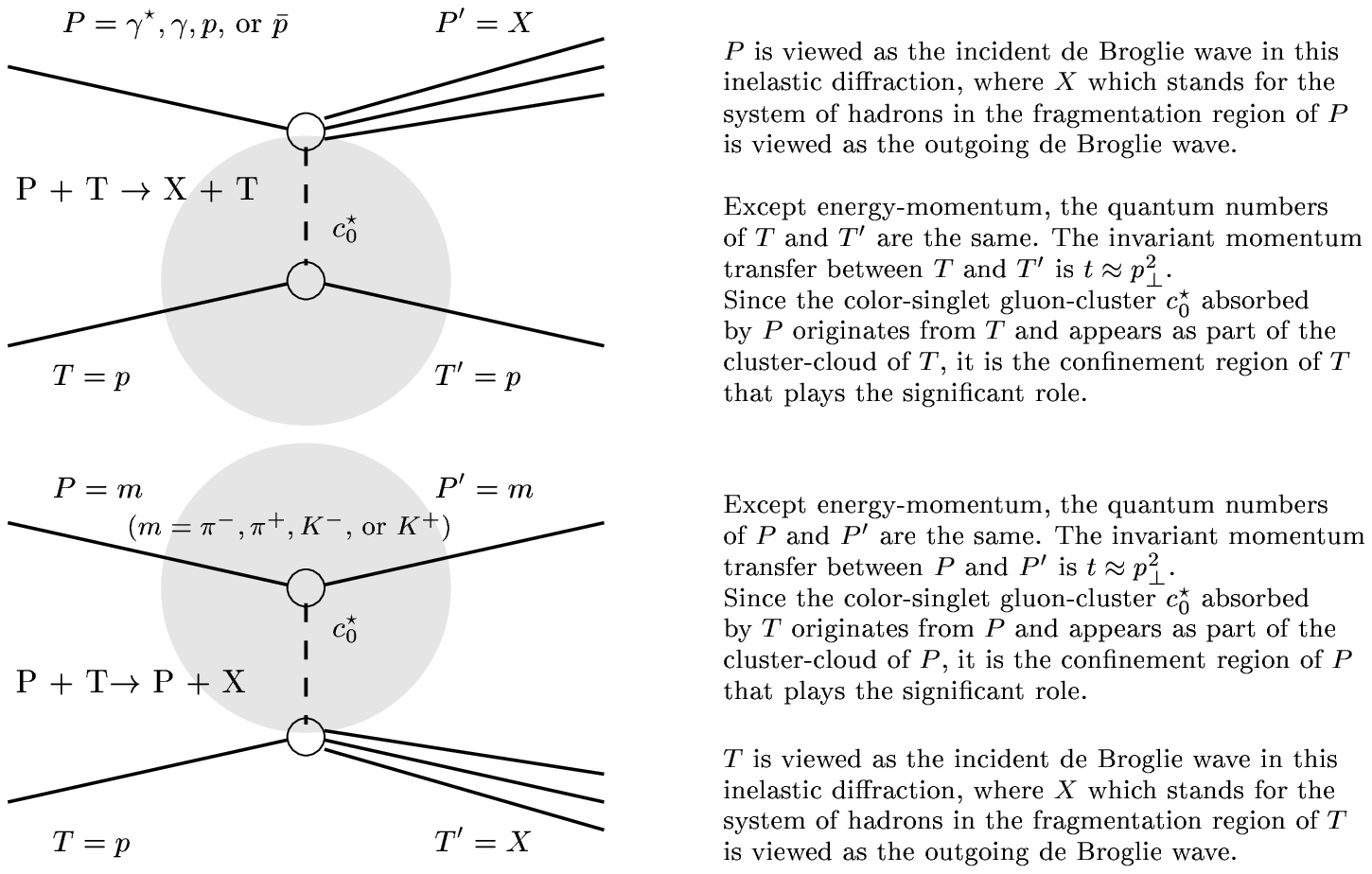}
\vspace*{1cm}
\caption{Inelastic diffraction in high-energy collision processes
mentioned in Eqs.(\ref{eq3}) and (\ref{eq4}), and those in
Eqs.(\ref{eq8}) and (\ref{eq9}),
viewed in the c.m.s. frame of the projectile ($P$) and the target ($T$). The
system of hadrons in the final state is a consequence of the (compared 
to elastic diffraction) relatively large transfer of
energy-momentum. The system is viewed as the product of $c_0^\star$
and the incident wave.}
\end{figure*}

\begin{figure*}
\hspace*{2cm}
\psfig{figure=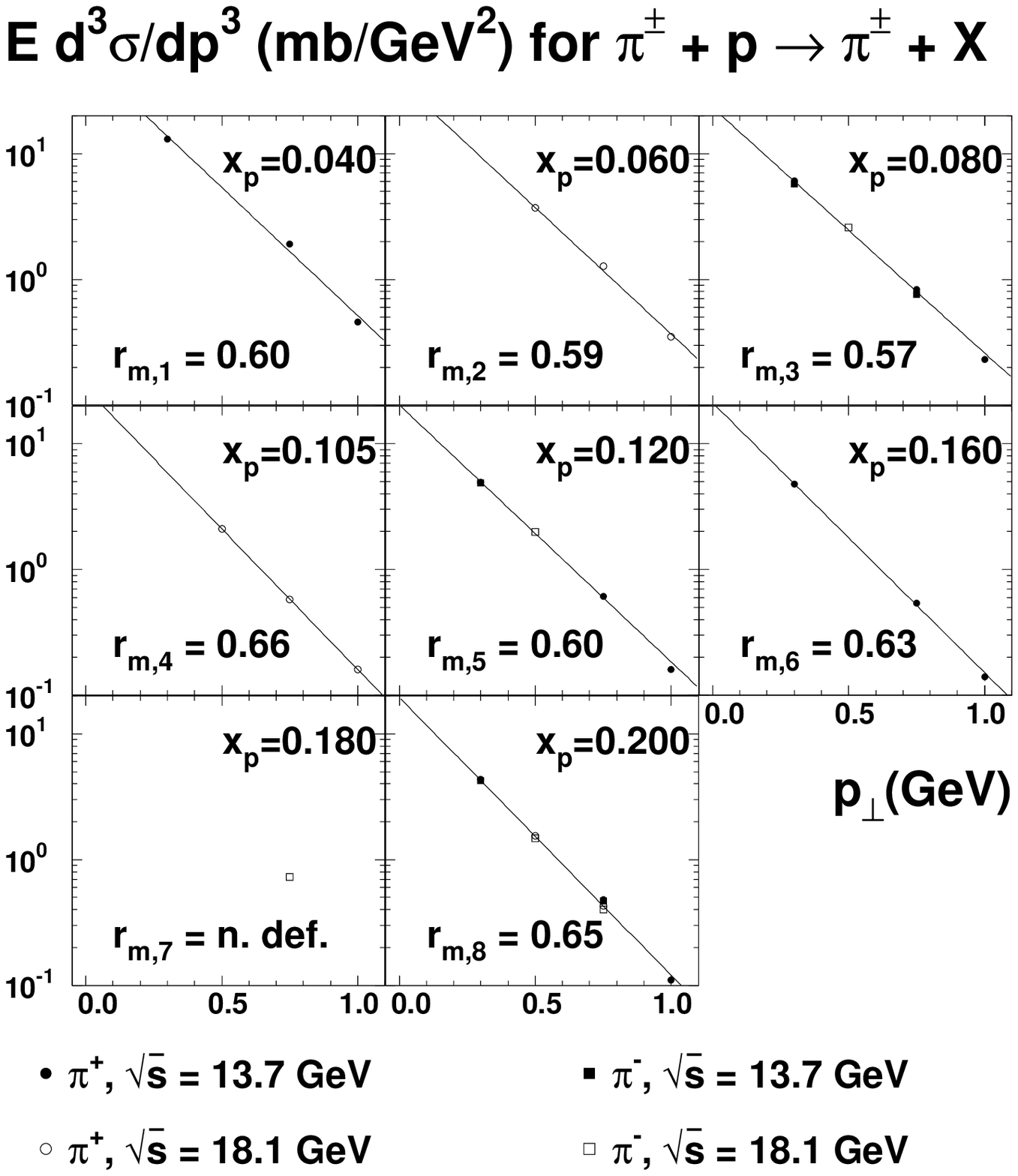,width=9cm}
\caption{Double differential cross section for diffractive inelastic 
$\pi^\pm +p\rightarrow \pi^\pm + X$ scattering as a function of the
transverse momentum of the scattered meson at given $x_P=M_x^2/s$. The 
data are taken from Ref.[\ref{diff-data}]. The lines are the result
of a fit to the data. The pion radius obtained by using the method
described in the text is $r_{\pi^\pm}= 0.62 \pm 0.04 \,\, \mbox{fm}$.}
\end{figure*}

\begin{figure*}
\hspace*{2cm}
\psfig{figure=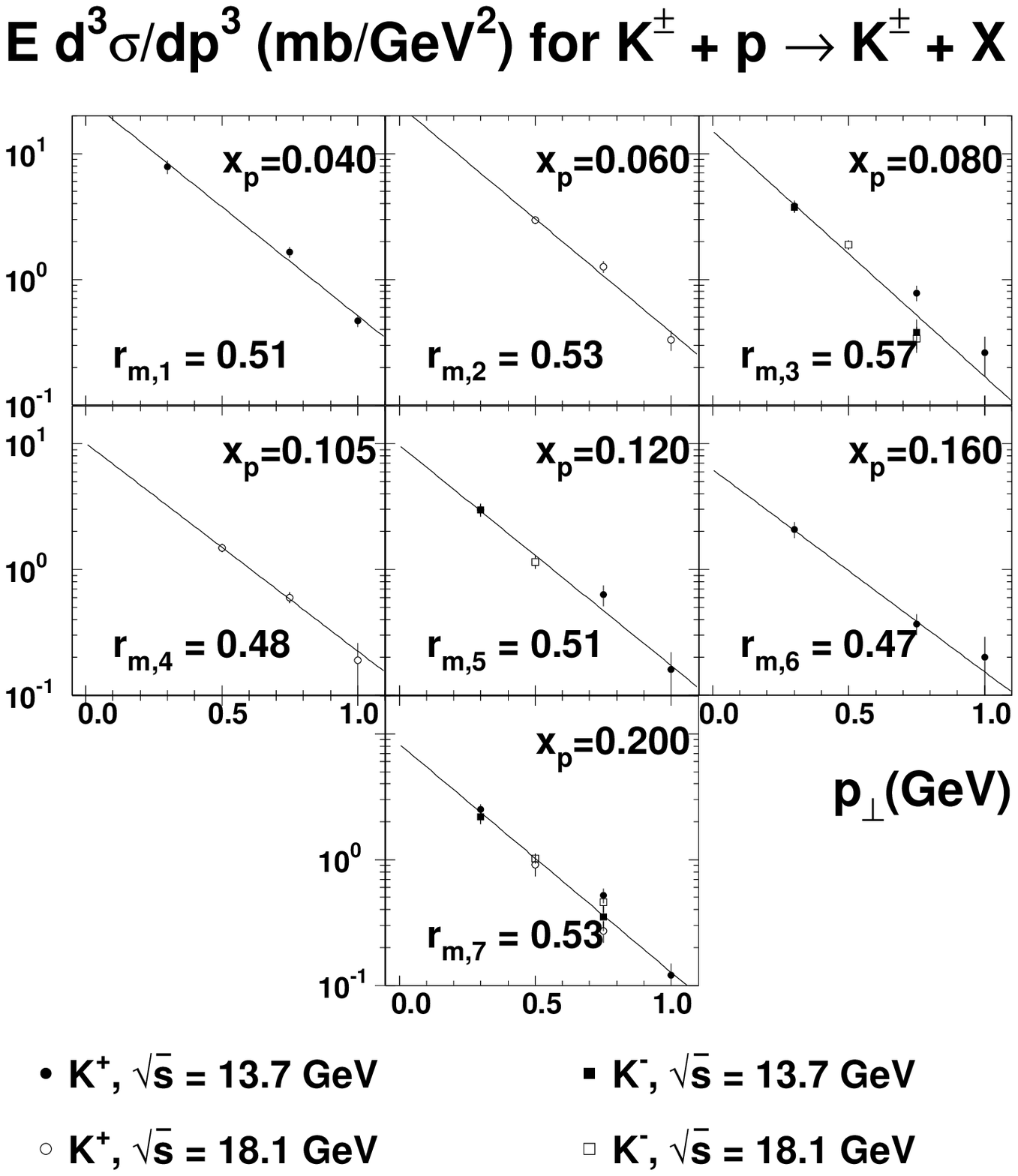,width=9cm}
\caption{Double differential cross section for diffractive inelastic 
$K^\pm +p\rightarrow K^\pm + X$ scattering as a function of the
transverse momentum of the scattered meson at given $x_P=M_x^2/s$. The 
data are taken from Ref.[\ref{diff-data}]. The lines are the result
of a fit to the data. The kaon radius obtained by using the method
described in the text is $r_{K^\pm}= 0.53 \pm 0.03 \,\, \mbox{fm}$.}
\end{figure*}

\begin{figure*}
\hspace*{4cm}
\psfig{figure=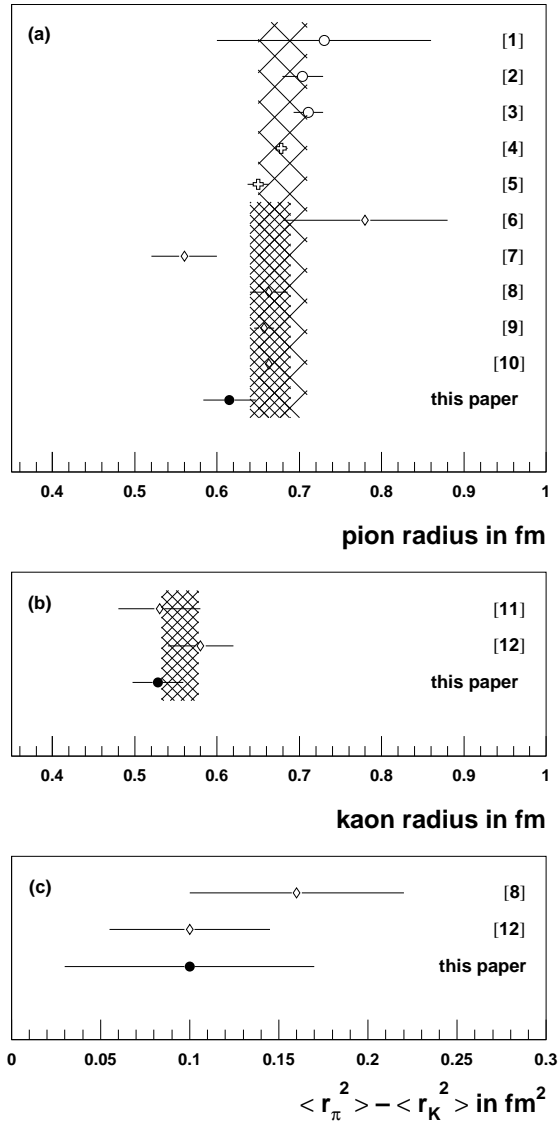,width=8cm}
\caption{\protect Experimental values for (a) pion and
(b) kaon radius, respectively [\ref{electro1}-\ref{direct7}].
The average value including the error of all the measurements 
given in Refs.[\ref{electro1}-\ref{direct7}] 
is indicated by the hatched
area. The more dense hatch pattern indicates the corresponding average
including error obtained from direct $\pi^- e^- \rightarrow \pi^- e^-$
scattering experiments [\ref{direct1}-\ref{direct5}] and 
direct $K^- e^- \rightarrow K^- e^-$ scattering
experiments [\ref{direct6}-\ref{direct7}]. 
(c) Experimental values for the difference of the squared radii of the
pion and the kaon [\ref{direct3},\ref{direct7}].
The solid points in (a), (b), and (c) 
are the results obtained in this paper, where
the data given in Ref.[\ref{diff-data}] are used.}
\end{figure*}

\end{document}